\documentclass{aa}  

\usepackage{graphicx}
\usepackage{txfonts}
\begin{document}

   \title{Frontier Fields: Combining \textit{HST}, \textit{VLT},  and \textit{Spitzer} data to explore the $z$$\sim$8 Universe behind the lensing cluster MACS0416$-$2403}


   \author{N. Laporte
          \inst{1}
          \and
          A. Streblyanska \inst{2,3}
          \and 
          S. Kim \inst{1}
          \and
          R. Pell\'o \inst{4}
          \and
          F. E. Bauer \inst{1,6,7}
           \and
          D. Bina \inst{4}
          \and
          G. Brammer \inst{8}
          \and
          M. A. De Leo \inst{5}
           \and
           L. Infante \inst{1,9}
          \and 
          I. P\'erez-Fournon \inst{2,3}
       }

   \institute{Instituto de Astrof\'{\i}sica, Facultad de F\'{i}sica, Pontificia Universidad Cat\'{o}lica de Chile, 306, Santiago 22, Chile \\
              \email{nlaporte@astro.puc.cl, fbauer@astro.puc.cl, skim@astro.puc.cl,linfante@astro.puc.cl}
             \and
             Instituto de Astrof\'{\i}sica de Canarias (IAC), E-38200 La Laguna, Tenerife, Spain.\\
             \email{alina@iac.es, ipf@iac.es}
             \and
             Departamento de Astrof\'{\i}sica, Universidad de La Laguna (ULL), E-38205 La Laguna, Tenerife, Spain
             \and
             IRAP, CNRS - 14 Avenue Edouard Belin - F-31400 Toulouse, France \\
             \email{roser.pello@irap.omp.eu, dbina@irap.omp.eu}
             \and
             Instituto de Astronom\'ia, Universidad Nacional Aut\'onoma de M\'exico, Mexico\\
             \email{deleo@astro.unam.mx}
             \and
             Millennium Institute of Astrophysics, Santiago, Chile
             \and
             Space Science Institute, 4750 Walnut Street, Suite 205, Boulder, Colorado 80301
             \and 
             Space Telescope Science Institute, 3700 San Martin Drive, Baltimore, MD 21218, USA
             \and
             Centro de Astroingenieria, Pontificia Universidad Cat\'{o}lica de Chile, Vicuna Mackenna 4860, Santiago, Chile 
             }

   \date{Received 23 September 2014; accepted 04 December 2014}

 
  \abstract
   {The {\it Hubble Space Telescope} ({\it HST}) Frontier Fields (HFFs) project started at the end of 2013 with the aim of providing extremely deep images of six massive galaxy clusters. One of the main goals of this program is to push several telescopes to their limits to provide the best current view of the earliest stages of the Universe. The analysis of the initial data has already demonstrated the huge capabilities of the program.}
   {We present a detailed analysis of  $z$$\sim$8 objects behind the HFFs lensing cluster, MACS0416-2403, combining 0.3-1.6 $\mu$m imaging from {\it HST}, ground-based $K_s$ imaging from VLT HAWK-I, and 3.6 $\mu$m and 4.5 $\mu$m \textit{Spitzer Space Telescope}. The images probe to 5$\sigma\ $ depths of $\approx$29 AB for {\it HST}, 25.6 AB for HAWK-I, and $\approx$0.310 and 0.391 $\mu$Jy at 3.6 and 4.5 $\mu$m, respectively. With these datasets, we assess the photometric properties of $z$$\sim$8 galaxies in this field, as well as their distribution in luminosity, to unprecedented sensitivity.}
   {We applied the classical Lyman break (LB) technique, which combines non detection criteria in bands blueward of the Lyman break at $z$$\sim$8 and color-selection in bands redward of the break. To avoid contamination by mid-$z$ interlopers, we required a strong break between optical and near-infrared data. We determined the photometric properties of $z$$\sim$8 selected candidates using spectral energy distribution (SED)-fitting with standard library templates. The luminosity function at $z$$\sim$8 is computed using a monte-carlo (MC) method taking advantage of the SED-fitting results. A piece of cautionary information is gleaned from new deep optical photometry of a previously identified $z$$\sim$8 galaxy in this cluster, which is now firmly detected as a mid-$z$ interloper with a strong $\approx1.5$ mag Balmer break (between F606W and F125W). Using the SED of this interloper, we estimated the contamination rate of our MACS0416$-$2403 sample, and that of previous samples in Abell 2744 that were based on HFF data, we highlight the dangers of pushing the LB technique too close to the photometry limits. }
   {Our selection reliably recovers four  objects with m$_{F160W}$ ranging from 26.0 to 27.9 AB that are located in modest-amplification regions ($\mu<$2.4). Two of the objects display a secondary break between the IRAC 3.6 $\mu$m and 4.5 $\mu$m bands, which could be associated to the Balmer break or emission lines at $z$$\sim$8. The SED-fitting analysis suggests that all of these objects favor high-$z$ solutions with no reliable secondary solutions. The candidates generally have star formation rates around $\sim$10 M$_{\odot}$/yr and sizes ranging from 0.2 to 0.5 kpc, which agrees well with previous observations and expectations for objects in the early Universe. The sample size and luminosity distribution are consistent with previous findings. }
   {}

\titlerunning{Frontier Fields : $z\sim$8 galaxies behind MACS0416}
\authorrunning{N. Laporte et al.}
   \maketitle
%

\section{Introduction}

Observations probing the edges of the Universe are among the most intriguing challenges of the coming decade, particularly with respect to detecting 
the first galaxies at $z$$>$12 (\citealt{2011ARA&A..49..373B}, \citealt{2011ApJ...740...13Z}). Several telescopes and instruments are under development for which the key objectives are these topics, such as the future E-ELT\footnote{E-ELT website : www.eso.org/public/teles-instr/e-elt/} \citep{2010SPIE.7735E..2DC}, NIRSPEC/JWST\footnote{JWST website : www.jwst.nasa.gov} \citep{2004SPIE.5487..653W}, and MOONS/VLT \citep{2012SPIE.8446E..0SC}. 

Many surveys have been completed to push the observational limits of the Universe even farther and to strongly increase the number of known very high-redshift sources ($z$$>$6). Ten years ago, only a dozen objects at $z$$>6$ had been discovered (\citealt{2004ApJ...616L..79B}; \citealt{2004ApJ...607..697K}), none above $z$$>$7.5. To date, the number of $z$$\sim$6, $z$$\sim$7, and $z$$\ge$8 galaxies selected in deep surveys count in the thousands (e.g., \citealt{2014arXiv1403.3938L}; \citealt{2014arXiv1403.4295B}; \citealt{2014MNRAS.438.1417M}; \citealt{2013AJ....145....4W}), several hundreds (e.g., \citealt{2011ApJ...737...90B}; \citealt{2009ApJ...706.1136O}; \citealt{2013ApJ...778..102O}; \citealt{2013ApJ...768...56T}) and about one hundred (e.g., \citealt{2013ApJ...777L..19L}; \citealt{2014ApJ...786..108O}; \citealt{2014ApJ...786...57S}; \citealt{2012Natur.489..406Z}; \citealt{2012ApJ...746...55T}), respectively. Thanks to the ever-increasing numbers of objects, the evolution and properties of galaxies is relatively well constrained up to $z$$\sim$6, with many secure spectroscopic confirmations (e.g., \citealt{2013ApJ...772...99J}; \citealt{2013AJ....145....4W}; \citealt{2014arXiv1408.3649S}; \citealt{2013MNRAS.429..302C}; \citealt{2012ApJ...744..179S}; \citealt{2014arXiv1403.3938L}). Beyond $z\sim$6, however, spectroscopic follow-up remains extremely challenging as a result of the decreasing mean brightness of these objects \citep{2013Natur.502..524F} and the nature of extreme mid-$z$ interlopers that may contaminate the high-$z$ sample  \citep{2012MNRAS.425L..19H}.

Several theoretical studies have demonstrated the merit of combining lensing fields and large deep blank fields to search for high-$z$ galaxies over a wide range of luminosities \citep{2010A&A...509A.105M}. The need of combining two kinds of fields has been confirmed by most of the current surveys, such as the \textit{Hubble Ultra Deep Field} \citep{2006AJ....132.1729B} and \textit{CLASH} \citep{2012ApJS..199...25P} carried out with the \textit{Hubble Space Telescope} ({\it HST}). The new flagship program of {\it HST}, the Frontier Fields (hereafter, HFFs) started at the end of 2013, and employs deep observations combined with gravitational lensing to observe six massive galaxy clusters along with six deep blank fields. Recent papers have shown that ultra faint galaxies ($m_{F160W}\sim$31-32 AB) will be detected in HFF data (\citealt{2014MNRAS.443L..20Y}, \citealt{2014MNRAS.444..268R}), and that the number of $z$$\ge$8 objects expected in these data is $\ga$130 \citep{2014arXiv1405.0011C}.

The data analysis of the first cluster Abell 2744 and its parallel field has already demonstrated the capabilities of this new legacy survey. A dozen $z$$\sim$8 objects have already been published (\citealt{2014ApJ...786...60A},b; \citealt{2014arXiv1405.0011C}; \citealt{2014arXiv1408.6903I}; \citealt{2014A&A...562L...8L}; \citealt{2014arXiv1402.6743Z}), and a lensed $z$$\sim$10 multiply imaged object has been detected close to the cluster core (\citealt{2014arXiv1407.3769Z}; \citealt{2014arXiv1409.1228O}). Observations of a second cluster, MACS0416$-$2403 ($z$=0.397), were just completed in September 2014. 

In this paper we describe the search for $z$$\sim$8 galaxies behind this second lensing cluster observed in the framework of the HFF combining {\it HST}, {\it VLT}, and {\it Spitzer} images. In Sect. \ref{sec.data_properties} we present the data we used, while in Sects.\ref{sec.highz_selection} and \ref{sec.phot_sample} we give details on the method we employed and the objects we included in our final sample. We discuss the properties of the selected sample in sect. \ref{sec.properties} and the contamination rate in this and other samples in sect. \ref{sec.contaminants}. The luminosity function (LF) at $z$$\sim$8 is computed in sect. \ref{sec.LF}, and a discussion including comparison with other samples can be found in sect. \ref{sec.discussion}. Throughout this paper, we adopt a $\Lambda$CDM cosmology with H$_0$=70 km s$^{-1}$ Mpc$^{-1}$, $\Omega_M$=0.3 and $\Omega_{\Lambda}$=0.7. Magnitudes are quoted in the AB system \citep{1983ApJ...266..713O}.

\section{Data properties}
\label{sec.data_properties}
\subsection{\textit{HST} images}
The HFF program combines data from the \textit{Advanced Camera for Survey} (ACS, \citealt{2005PASP..117.1049S}) in F435W, F606W, and F814W and the \textit{Wide Field Camera 3} (WFC3, \citealt{2008SPIE.7010E..1EK}) in F105W, F125W, F140W, and F160W. We used images reduced by the \textit{Space Telescope Science Institute} (STScI),refer to the STScI website for a description of the data reduction\footnote{www.stsci.edu/hst/campaigns/frontier-fields/}. The ACS final images include data from several {\it HST} programs observed in mid-2012 (ID 12459, PI: M. Postman) and early 2014 (ID 13496, PI: J. Lotz and ID 13386, PI: S. Rodney), corresponding to 21, 13, and 50 orbits in the F435W, F606W, and F814W bands. The WFC3 images combine data from the same {\it HST} programs and consist of 24, 12, 10, and 24 orbits in the F105W, F125W, F140W, and F160W bands. We measured the image depths using 100s of empty 0\farcs2 radius apertures distributed all over the field. These data reach a depth of $\sim$29 mag. at 5$\sigma$.

\subsection{\textit{Spitzer} images}
The {\it Spitzer} Space Telescope is also involved in the HFF project, and the Spitzer Science Center (SSC) provided Basic Calibrated Data (cBCD hereafter) for all the selected clusters (PI: T. Soifer and P. Capak).  These data are automatically corrected by the pipeline for various artifacts such as mumbled, muxstripe, and pulldown. We processed, drizzled, and combined these frames and the associated mask into final mosaics using the standard SSC reduction software MOPEX. The sensitivity of Spitzer data was computed using a similar method as for the HST images, using 100s of empty 1.4'' apertures all over the field. The final dataset we used in this study has a 5$\sigma$ point-source sensitivity of 0.310 and 0.391 $\mu$Jy at 3.6 and 4.5 $\mu$m.

\subsection{\textit{VLT} image}
\label{sec.hawki}
To improve our spectral energy distribution (SED) constraints, we added deep $K_s$ imaging from the HAWK-I instrument at the \textit{VLT} \citep{2004SPIE.5492.1763P} to our \textit{HST} and \textit{Spitzer} data. The HAWK-I observations were made between November 2013 and February 2014 as part of the ESO program 092.A-0472 (PI: G. Brammer).  The raw HAWK-I images were processed using a custom pipeline, which was originally developed for the NEWFIRM Medium Band Survey (NMBS; \citealt{2011ApJ...735...86W} ) and later adapted for the ZFOURGE \citep{2014ApJ...787L..36S} and HAWK-I HFF  surveys (Brammer, in prep.).  The final reduced HAWK-I mosaic has excellent image quality ($0\farcs4$ FWHM) and reaches a 5$\sigma$ limiting magnitude of  $K_s = 26.3$ for point sources.  The $7^\prime\times7^\prime$ HAWKI field of view covers both the cluster and parallel \textit{HST} pointings.  

Properties of the dataset we used are summarized in Table \ref{data_properties}.


\begin{table}
\caption{Properties of the \textit{HST} and \textit{Spitzer} data .}             
\label{data_properties}      
\centering                          
\begin{tabular}{c | c c c c l}        
\hline\hline                 
Filter & $\lambda_{central}$ & $\Delta\lambda$ & t$_{exp}$ & m(5$\sigma$) & Instrument \\    
         &  [$\mu$m]  & [nm]                   &   [ks]      & [AB]                        &                  \\          
 \hline                        
 F435 W	& 0.431 & 72.9   &  54.5     &   28.4            & ACS \\
 F606W	& 0.589 & 156.5 &  33.5     &   29.0            & ACS \\
 F814W	& 0.811 & 165.7 &  129.9   &   28.7            & ACS \\
 F105W	& 1.050 &	300.0 &  67.3	 &   29.4		& WFC3 \\
 F125W	& 1.250 &	300.0 &  33.1	 &   29.2		& WFC3 \\
 F140W	& 1.400 & 400.0 &  27.6	 &   29.1		& WFC3 \\
 F160W	& 1.545 & 290.0 &  66.1	 &   29.1		& WFC3 \\
\hline
K$_s$	& 2.146 & 0.324 &  97.4     &   26.3            & HAWKI\\
\hline
 3.6	& 3.550 & 750.0   &	160.2 & 25.2 & IRAC \\
 4.5 	& 4.493 & 1015.0 &  180.0 & 24.9 & IRAC \\
\hline
\hline                                   
\end{tabular}
\tablefoot{Columns: (1) filter ID, (2) filter central wavelength, (3) filter FWHM, (4) exposure time, (5) depth at 5$\sigma$ in a 0\farcs2 radius aperture for {\it HST} data, 0\farcs45 radius aperture for \textit{HAWKI} data, and 1\farcs4 radius aperture for IRAC images, (6) instrument. }
\end{table}
%
%

\section{Selection of high-$z$ candidates}
\label{sec.highz_selection}
In the following, we briefly explain how we constructed our photometric catalogs and describe the method we used to select $z$$\sim$8 candidates. We conclude by discussing the completeness of this survey.

\subsection{Photometric catalogs}
We used SExtractor v2.19.5 \citep{1996A&AS..117..393B} in double-image mode to construct WFC3 catalogs, using F125W as detection image, and in single-image mode for the ACS catalogs to avoid any false detections at optical bands. The extraction parameters we used are similar to those presented in \citet{2014arXiv1408.6903I} and are defined to maximize the detection of faint objects and objects in crowded regions close to the cluster core: $DETECT\_MINAREA$ 6 pixels, $DEBLEND\_NTHRESH$ 16 and $DEBLEND\_MINCONT$ 0.0005. We then matched all of the catalogs using TOPCAT \citep{2005ASPC..347...29T}, allowing for a maximum error shift between the position of object in WFC3 data and ACS data of 0\farcs03, with the understanding that the {\it HST} data used have an astrometric rms of 3-4mas\footnote{see README file provided by STScI associated with each data release}. The use of two different methods to build ACS and WFC3 catalogs is motivated by the nature of the field. Indeed, close to the cluster core, the noise increases strongly and the dual-image mode of SExtractor could then find "false" detections. We quantified this effect by running SExtractor in dual- and simple- image mode on all the three ACS images. As a result, we found after visual inspection that $\approx$9\% of the objects detected in the double-image mode catalog are not detected in simple-image mode.

The final catalog contains $\sim$30,000 detections distributed over the $\sim$2'$\times$2' WFC3 field of view of MACS0416$-$2403.

\subsection{Selection criteria}
\label{selection_criteria}
We applied the Lyman break (LB) technique \citep{1996ApJ...462L..17S}, one of the most popular methods aiming to select very high-redshift objects (e.g., \citealt{2014arXiv1403.4295B}; \citealt{2013ApJ...773...75O}; \citealt{2012Natur.489..406Z}) . It combines non detection criteria in bands blueward of the LB and color selection in bands redward of the break. To select objects at $z$$\geq$7.5, we required a 5$\sigma$ detection in F125W and F140W and less than a 2$\sigma$ detection in all of the ACS bands. Moreover, to avoid selecting extreme low-$z$ interlopers, we required a break between F814W and F125W of $>$2.5 mag (see Section \ref{sec.contaminants} and \citealt{2012MNRAS.425L..19H}); this requires F125W$<$28.2. 

We psf-matched the WFC3 images to the psf of the F160W image using the Tiny-Tim model \citep{2011SPIE.8127E..0JK} and used the color-criteria defined in \citet{2014ApJ...786...60A} on these data: 
\begin{multline*}
F105W - F125W > 0.5 \\
F105W - F125W > 0.4 + 1.6\times(F125W-F140W) \\
F125W - F140W < 0.5. \\
\end{multline*}
\noindent Error bars were estimated using source-free apertures around the object. We performed visual inspection to make sure that false detections were removed from the sample, such as groups of pixels in the close neighborhood of bright galaxies. These regions are indeed masked out from the statistical analysis, but spurious sources could survive in the surrounding area. This kind of false detection is difficult to avoid given the DETECT\_MINAREA used for extraction, which is optimized for detecting compact faint sources. In addition, visual inspection also allowed us to confirm (i.e., recompute) the photometry for real sources located in crowded areas. The general process adopted here is similar to \citet{2014arXiv1410.5439F}. Among the 473 detections that followed the selection criteria, we retained four (4) sources as good $z$$\sim$8 candidates.The large number of detections that need to be inspected visually is due to the SExtractor parameters used to select point-source-like objects. Because the high-$z$ nature of an object is given by the shape of its SED in the wavelength covered by HST data, we use in Section \ref{sec.phot_sample} data at longer wavelengths to better constrain the SED and identify more probable low-$z$ interlopers.. 
%
%

\subsection{Completeness of the survey}
The completeness of a photometric survey can be separated into two parts:  the completeness of the method used to extract objects, and the incompleteness implied by the use of a color-selection, as follows:
\begin{equation}
C_{tot}(m,z) = C_{ext}(m)\times C_{color}(m,z).
\label{eq.completeness}
\end{equation}

\begin{figure}
   \centering
           \includegraphics[width=8cm]{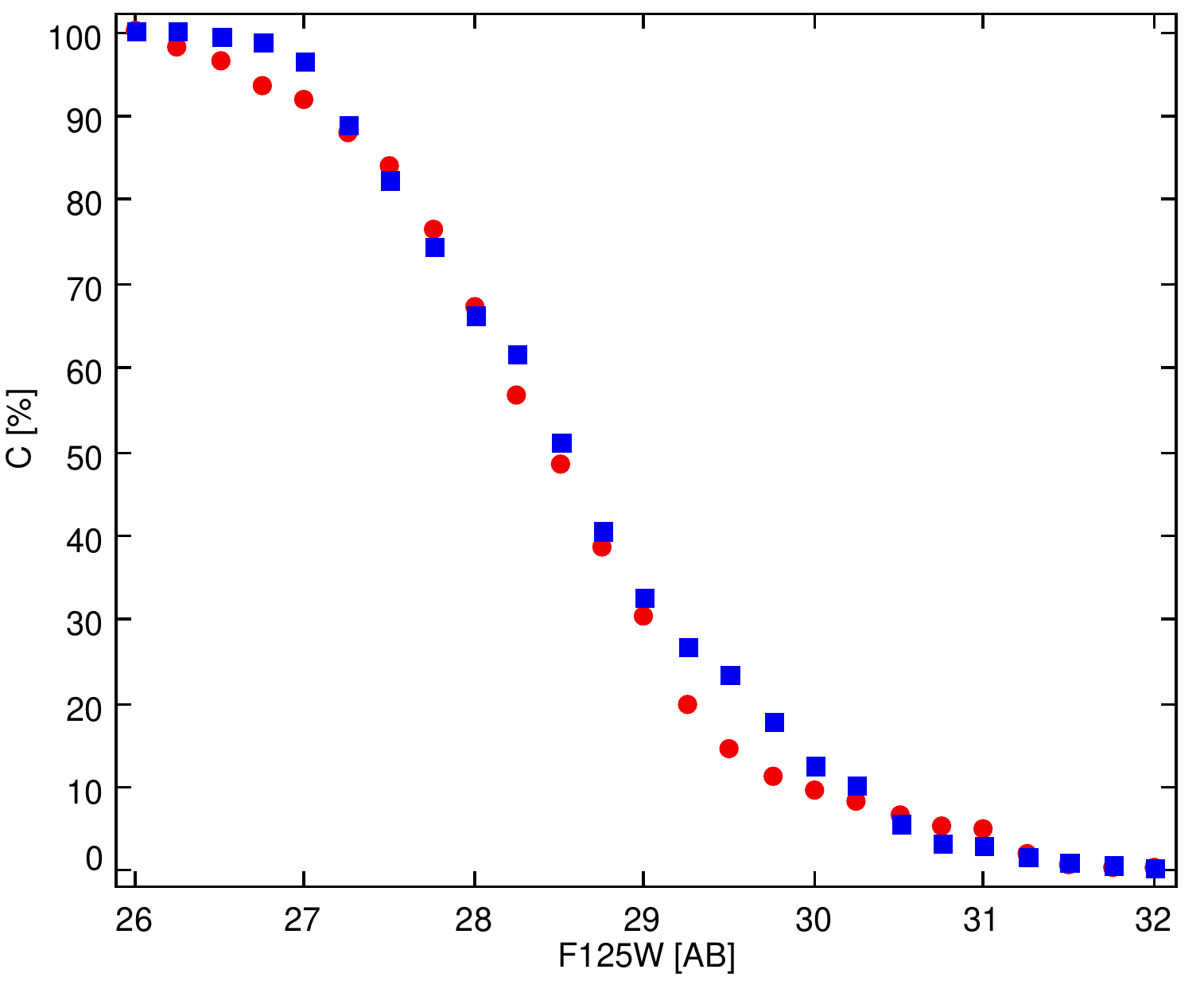}
      \caption{\label{fig.detection_completness}Completeness of the extraction method used to select $z$$\sim$8 objects. We simulated 1000 galaxies per bin of 0.25 magnitude and compared the number of extracted sources with the number of added objects. Blue dots shows the completeness computed far from the cluster core, the red dots are computed close to the center, in a 1.8'$\times$0.5' rectangular region centered on the cluster core.}
   \end{figure}

 To estimate the completeness of the extraction method, $C_{ext}(m)$, we added 1000 artificial point-source-like objects per bin of 0.25 magnitude to the detection image and applied the same extraction parameters as we used to build previous catalogs. The ratio between the number of objects extracted and the number of objects added to the data gives us the first part of the completeness function. Moreover, to show the influence of the cluster core on the completeness, we performed two simulations, one by adding objects only close to the cluster core in a 1.8'$\times$0.5' rectangular region centered on the cluster core, and the other one considering only regions far from the cluster core, outside of the previous rectangular region. Figure \ref{fig.detection_completness} displays the results of these simulations and shows that the cluster core has a negligible effect on the completeness.
 
\begin{figure}
   \centering
           \includegraphics[width=10cm]{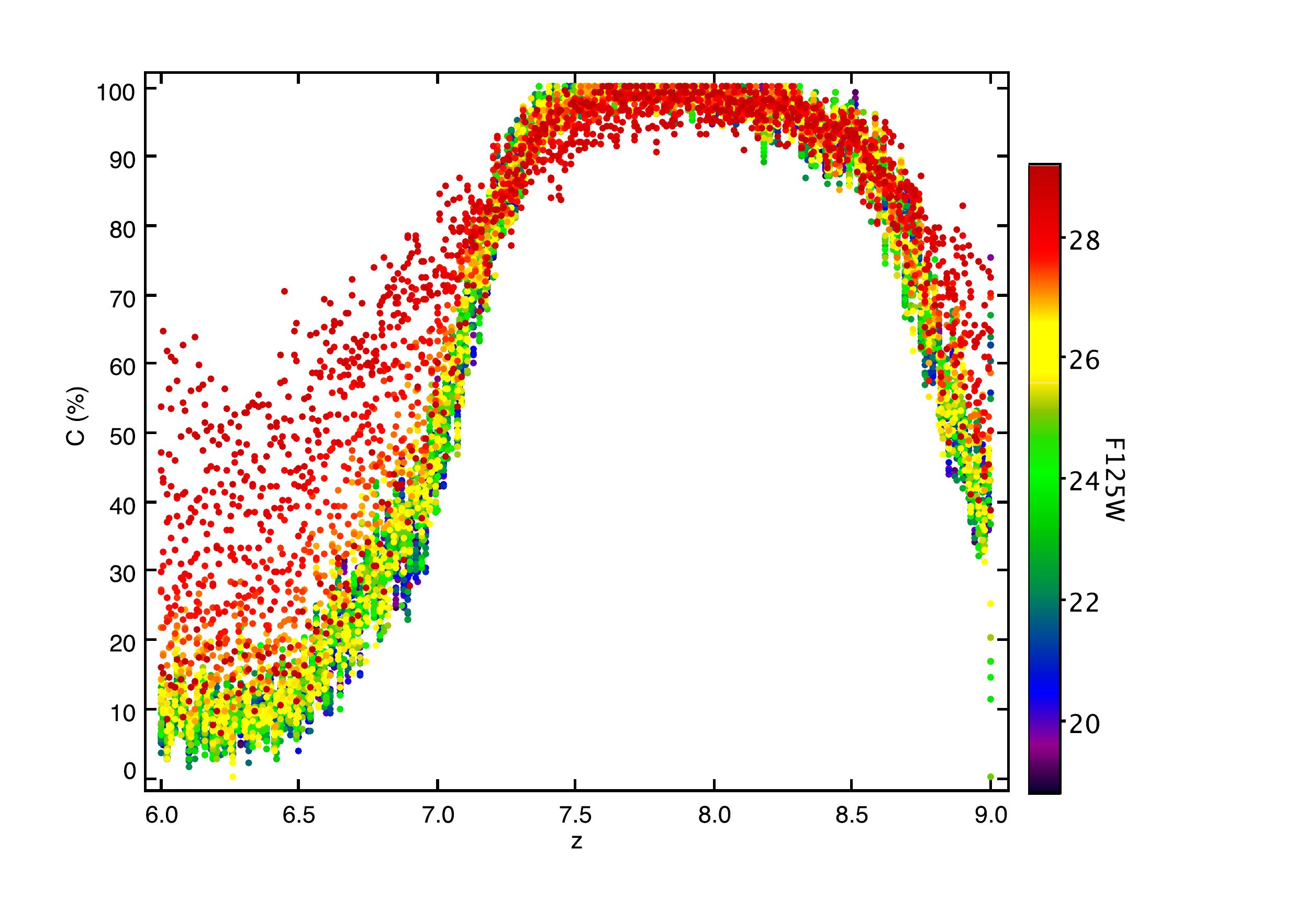}
      \caption{\label{fig.color_completness}Completeness of the color criteria used to select $z$$\sim$8 objects as a function of redshift and magnitude on the detection image. We simulated 100,000 galaxies per bin of 0.25 magnitude with a redshift ranging from 0 to 10 from standard templates (\citealt{2003MNRAS.344.1000B}, \citealt{1980ApJS...43..393C}, \citealt{1996ApJ...467...38K}, \citealt{2007ApJ...663...81P} and \citealt{1998ApJ...509..103S}).}
   \end{figure}

 The second term of Equation \ref{eq.completeness}, $C_{color}(m,z)$, was obtained by simulating galaxy SEDs from standard templates and applying the color criteria we used to select high-$z$ galaxies. We used templates from \citet{2003MNRAS.344.1000B}, \citet{1980ApJS...43..393C}, \citet{1996ApJ...467...38K}, \citet{2007ApJ...663...81P}, and \citet{1998ApJ...509..103S}  and simulated 100,000 objects per bin of 0.25 magnitude in a redshift interval ranging from 0.0 to 10.00. We used filter transmissions and noise measured on each image as a function of magnitude to reproduce the photometric quality of real data.  We then applied the color criteria defined in Sect. \ref{selection_criteria} and defined the completeness as the ratio between the number of simulated and selected galaxies (Figure \ref{fig.color_completness}). 
%
%

%
%

\section{Photometric sample}
\label{sec.phot_sample}
We selected four objects, named from the brightest to the faintest MACS0416$\_$Y1 to $\_$Y4 hereafter, as good $z$$\sim$8 candidates with F125W magnitude ranging from 26.3 to 27.5. Two objects of the selected sample are in common with the $z$$\sim$8 sample selected using previously released {\it HST} data alone that were published by \citet{2014arXiv1405.0011C} (MACS0416$\_$Y2 = FFC2-1153-4532 and MACS0416$\_$Y3 = FFC2-1151-4540). To perform SED-fitting by combining photometry from several instruments, we need to estimate the total flux in each band. We obtained the total magnitudes of our candidates following \citet{2014A&A...562L...8L}, assuming that the total flux is given by the F160W SExtractor MAG\_AUTO, and corrected for aperture effects deduced from Kron photometry \citep{2013Natur.502..524F}.

Of the selected sample,the two objects MACS0416$\_$Y1 and MACS0416$\_$Y3 are also detected in the deep {\it Spitzer} data. Both show a non detection at 3.6 $\mu$m with a clear detection at 4.5 $\mu$m, and thus display a SED similar to previously published $z$$\sim$8 objects (\citealt{2014A&A...562L...8L}, \citealt{2013Natur.502..524F}). This break in the {\it Spitzer} data can be ascribed to the Balmer break and additional contamination by [OIII] and H$\beta$ emission lines at $z$$\sim$8, which strengthens the high-$z$ solutions for these objects (see Figure 1 of \citealt{2014ApJ...784...58S}).  

Because of the high number density of objects in the {\it Spitzer} bands, we extracted the photometry of the two sources detected at 4.5$\mu$m using the galaxy-fitting program GALFIT \citep{2002AJ....124..266P}. We followed the procedure described in \citet{2013Natur.502..524F} to fit and subtract all nearby sources around the objects. We then measured the photometry in a residual (contamination-free) image in a 1\farcs4 radius aperture and applied standard aperture correction \footnote{see the IRAC instrument handbook for more details}. 

We added one more constraint between 1.6 and 3.5 $\mu$m using the HAWK-I $K_s$ image we described in Section \ref{sec.hawki} \footnote{ We also used the public $K_s$ image obtained with GeMS \citep{2008SPIE.7015E..0XB} installed on Gemini-South, but this image is not deep enough to add a robust constraint on the SED of our objects \citep{2014arXiv1409.1820S}.}. Two objects from our sample are clearly detected in that deep image (Y1 and Y3). For the non detected object, we measured the flux at the position of our $z$$\sim$8 candidates, and, to be more conservative, we set as an upper limit 2$\sigma$. Table \ref{tab.z8_phot} and Fig. \ref{fig.z8_stamps} display the photometry and image stamps of the four objects retained as good $z$$\sim$8 candidates, and Fig. \ref{fig.FoV} displays the position of these sources over the WFC3 field of view.


\begin{table*}
\caption{Photometric table of $z$$\sim$8 candidates selected behind MACS0416}             
\label{tab.z8_phot}      
\centering                          
\begin{tabular}{c | c c | c c c c c c c c c l}        
\hline\hline               
ID			 & RA$_{J2000}$   & DEC$_{J2000}$  &	$Y_{105}$ & $J_{125}$ 	& $JH_{140}$ & $H_{160}$ &  $K_s$ & 3.6 & 4.5 \\    
         		& [deg] &	[deg] &		&		&		&             &                  &	 &        \\          
 \hline                        
		 		&			&			&			&			&			&      		        &			&            		&       		       \\
MACS0416\_Y1 	&      64.03917 	& -24.09318	&	$>$29.2	&	26.29	&	26.06	& 	26.05	&	26.40	&   $>$25.70     &  24.93		         \\
				&			&		   	&			&$\pm$0.03	&$\pm$0.02	&$\pm$0.02	&$\pm$0.20	&           		&  $\pm$0.22	       \\ 
MACS0416\_Y2	&	64.04799 	& -24.08167	&	28.48	&	26.51	&      26.55        &     26.51		&      $>$26.48	&   $>$25.70     & $>$25.50		   \\
 		  		&			&			&   $\pm$0.16	&$\pm$0.05	&$\pm$0.03	&$\pm$0.02	&			&            		&        		       \\ 
MACS0416\_Y3	&	64.04804 	& -24.08143	&	$>$29.1	&	27.14	&      26.67	&     26.42         &     26.52   	&$>$25.70     &   25.34   	  \\
				&			&			&			&$\pm$0.07	&$\pm$0.04	&$\pm$0.03	&$\pm$0.22	&           		 &   $\pm$0.32	       \\ 
MACS0416\_Y4	&	64.03756	& -24.08810	&	$>$29.1	&	27.44	&     27.45	        &     27.91         &      $>$26.4	 &   $>$25.70     &  $>$25.50  	   \\
 				&			&			&			&$\pm$0.13	&$\pm$0.07	&$\pm$0.11	&			&            		&       		       \\  
\hline
\hline                                   
\end{tabular}
\tablefoot{{\it HST} magnitudes given in this table are total magnitudes estimated from SExtractor MAG\_AUTO magnitude and corrector for aperture effects. {\it Spitzer} photometry points are measured on contamination-free images. Upper limits are given at 2$\sigma$ in 0\farcs4, 0\farcs45, and 1\farcs4 radius aperture at the position of the source, in WFC3, VLT, and IRAC images. \\
}
\end{table*}
%
%

\section{Properties of the $z$$\sim$8 candidates}
\label{sec.properties}
The photometric properties of our $z$$\sim$8 candidates were deduced using an SED-fitting approach and were corrected for magnification by the cluster.

\subsection{Magnification}
\label{sec.magnification}
Five teams have provided amplification maps for the six HFF clusters, using different methods and assumptions on the mass models (\citealt{2009ApJ...706.1201B};  \citealt{2014MNRAS.444..268R}; \citealt{2011MNRAS.417..333M}; \citealt{2013ApJ...762L..30Z}; \citealt{2014arXiv1405.0222J}; \citealt{2014MNRAS.439.2651M}). \citet{2014MNRAS.443.1549J} have used the Frontier Fields data to search for multiple-image objects in MACS0416$-$2403 data. They used the best 57 multiple-image galaxies and the  software package \textit{Lenstool} \footnote{website : http://projects.lam.fr/repos/lenstool/wiki} to provide the highest precision mass model for this cluster to date. For this reason, we use the amplification map provided by the CATS group in the following. All of the candidates presented in this paper lie in regions with only modest amplification ($\mu\sim$1.5 to 1.9).

\subsection{SED-fitting}
\label{sec.sed-fitting}
We used the public code Hyperz (v12.2, \citealt{2000A&A...363..476B}) to fit the SED of each object with a library of 14 templates: 8 evolutionary synthetic SEDs extracted from \citet{2003MNRAS.344.1000B}, with Chabrier IMF \citep{2003PASP..115..763C} and solar metallicity; a set of 4 empirical SEDs compiled by \citet{1980ApJS...43..393C}, and 2 starburst galaxies from the \citet{1996ApJ...467...38K} library. The flux was set to zero in the bands where an object was undetected, with an error bar corresponding to the limiting flux at the position of the candidate.

We first fitted the SED without any constraint on the redshift interval, allowing values of between $z$=0.0--12.0, and a large reddening interval ($A_v$=0.0--3.0, following \citealt{2000ApJ...533..682C}). For all of the objects in our sample, the best-fit SED is found at very high-redshift ($z$$>$8.0) with a small 1$\sigma$ confidence interval ($dz$$<$1), meaning a good estimation of the photo-$z$. For most of the objects, the best-fit SED is found to have negligible dust content ($A_v$=0), as expected for very high-redshift objects (e.g., \citealt{2010A&A...515A..73S}). We also inspected the redshift probability distribution, $P(z)$ to evaluate the probability that they are low-$z$  interlopers. Only one object displays a secondary solution, MACS0416\_Y3, but the integrated probability associated with this low-$z$ solution, P($z\sim$2)=0.2\%, is low.

\begin{figure*}
   \centering
           \includegraphics[width=18cm]{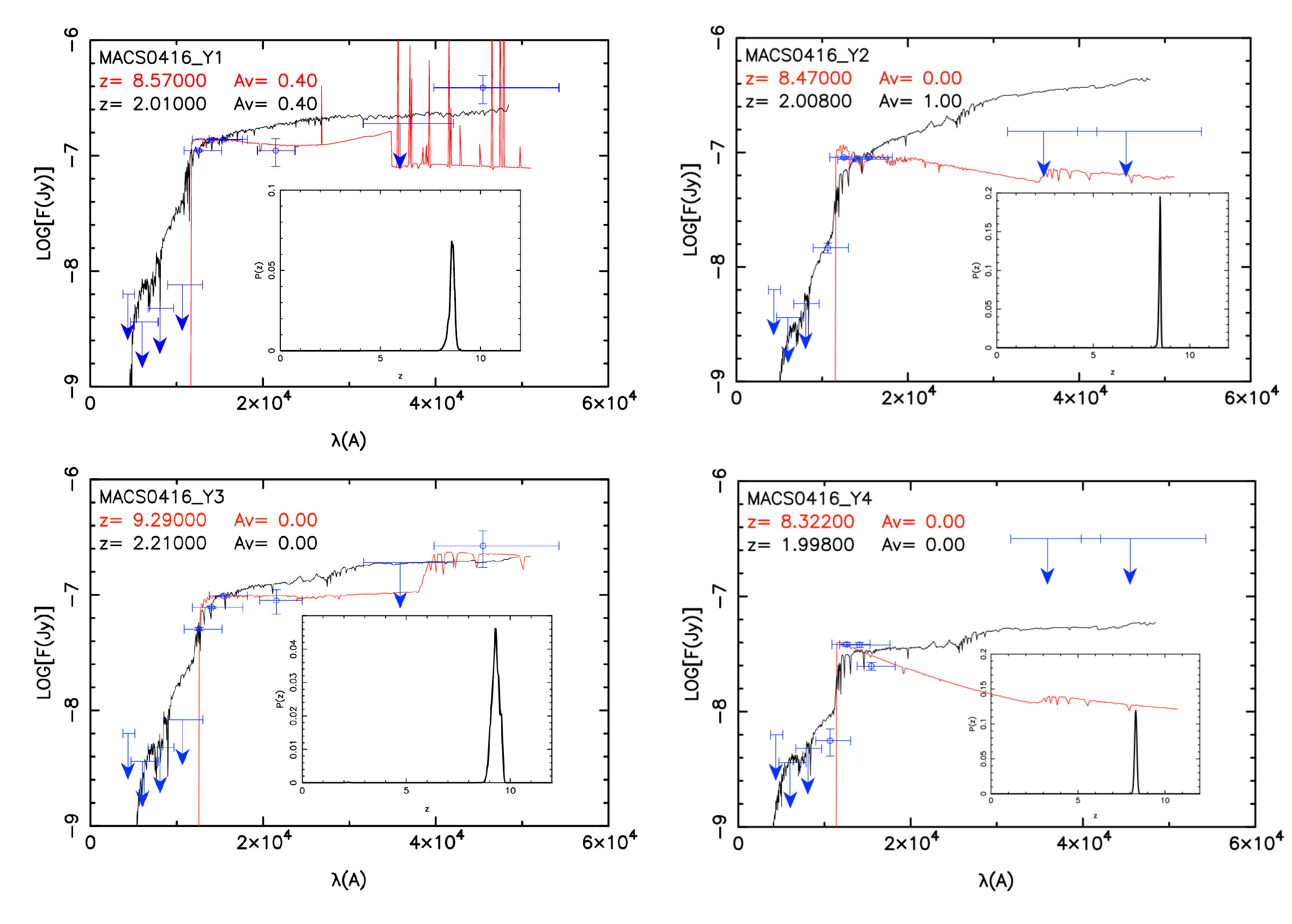}
      \caption{\label{fig.sed_fit} Best SED fit for the 4 $z$$\sim$8 candidates selected behind MACS0416. The red line displays the best SED fit found without any constraint on the redshift interval (0$<$$z$$<$12), the black line shows the best fit assuming a low-$z$ solution (0$<$$z$$<$3). Upper limits are 2$\sigma$ non detection. The redshift probability distribution (\textit{P(z)}) is also plotted as an inset of each fit.}
   \end{figure*}

To study the contribution of emission lines in the Spitzer photometry for MACS0416\_Y1 and \_Y3, we also used library templates built from \textit{Starburst99} templates \citep{1999ApJS..123....3L} adapted to include nebular emissions. The best SED fit is always found at high-$z$ for these two objects, even if the contribution of $z$$\sim$8 emission lines for MACS0416\_Y3 seems negligible (cf. Figure \ref{fig.sed_fit}). 

We repeated the SED fitting procedure for all of the objects using the same library of templates, but reducing the redshift interval (0.0$<z<$3.0). Indeed, recent spectroscopic studies have shown that most of the interlopers that contaminated $z$$\sim$8 samples are at $z$$\sim$2 (e.g., \citealt{2014arXiv1410.5558P}; \citealt{2013ApJ...765L...2B}; \citealt{2012MNRAS.425L..19H}).  For all objects, the best-fit SED has a significantly larger $\chi^2$ and for three of them requires higher dust reddening than in the previous case. Moreover, best SED fits are inconsistent either with the IRAC constraints or the ACS non detection.  


\begin{table*}
\caption{Photometric properties of $z$$\sim$8 candidates selected behind MACS0416}             
\label{tab.z8_prop}      
\centering                          
\begin{tabular}{c | c c c c | c c c | c c c| c  l}        
\hline\hline                 
ID			 	&$z_{phot}$	& $\chi^2_{red}$	& $A_v$	& $\Delta$z$$	&$z_{phot}^{low}$	& $\chi^2_{red}$	& $A_v$	& $L_{1500}$				& SFR				&	Size			&  $\mu$  \\    
         			& 			&				& [mag]	&		&				&            			 & [mag]   & $\times$10$^{40}$[erg/s]	& [M$_{\odot}$/yr]  		& [kpc]			&     \\          
 \hline                        
		 		&			&				&		&		&				&      		      		&            	&       					&       				&		\\
MACS0416\_Y1 	& 8.60		&	0.69			& 0.0		& 8.1 - 8.9& 2.01			& 5.46			& 0.40	&      12.3$^{+0.6}_{-1.0}$ 	&   12.9$^{+0.6}_{-1.0}$	& 0.44$\pm$ 0.03	&	1.5		\\
			(*)	& 8.57		&	0.64			& 0.40	& 8.1 - 8.9	&				&				&		&						&		&		\\
			 	& 			&				&		&		&				&      		      		&            	&       					&       	&		\\ 
MACS0416\_Y2	& 8.47		&	 1.09			& 0.0		& 8.3 - 8.6& 2.01			& 11.30			& 1.00	&       7.8$^{+0.1}_{-0.3}$ 		&  8.1$^{+0.2}_{-0.2}$     	& 0.31$\pm$ 0.05	&	 1.6		\\
 		  		&			&				&		&		&				&      		      		&            	&       					&       	&		\\
MACS0416\_Y3	& 9.35		&	0.29			& 0.40			& 8.8 - 9.7			& 2.21			& 1.49	& 0.00		& 23.9$^{+1.4}_{-0.6}$ 			& 25.2$^{+1.4}_{-0.9}$				& 0.31$\pm$ 0.04	&	1.6		\\
			(*)	&9.29		&	0.24			& 0.0 			& 8.8 - 9.7			&				&				&		&						&		&	\\
			 	& 			&				&		&		&				&      		      		&            	&       					&       	&		\\ 
MACS0416\_Y4	& 8.00		&	2.43			& 0.0		& 7.3	 - 8.3& 1.90			& 7.78 			& 0.00	& 2.1$^{+0.2}_{-0.3}$ 		& 2.3$^{+0.1}_{-0.3}$	& 0.19$\pm$ 0.03	&	1.9		\\
 				&			&				&		&		&				&      		      		&            	&       					&       	&		\\
\hline
\hline                                   
\end{tabular}
\tablefoot{Columns: (1) ID, (2, 3, 4, 5) photometric redshift, $\chi^2$, reddening and 1$\sigma$ confidence interval deduced from SED-fitting without any constraint on the redshift, (6, 7, 8)  photometric redshift, $\chi^2$ and reddening deduced from SED-fitting with a redshift interval set between 0.0 and 3.0, (9,10) luminosity at 1500\AA\ and Star formation rate computed from \citet{1998ARA&A..36..189K} and corrected for dust extinction and amplification,  the amplification factor (11) measured on maps provided by the CATS group \citep{2014MNRAS.444..268R}.  \\
(*) Best SED-fit results from the templates library including nebular emissions.
 }
\end{table*}
%
%

%
%

\subsection{Star formation rate and size}
\label{sec.SFR_size}
We computed the star formation rate (SFR) of our sample from the SED fits described in Section \ref{sec.sed-fitting}. From the best-fit SED we computed the luminosity at 1500\AA\ and corrected it for magnification and also reddening effects following the equations given in \citet{2000ApJ...533..682C}. We then computed the SFR from the \citet{1998ARA&A..36..189K} relationship between UV luminosity and SFR. Our candidates have an SFR $\leq$25 M$_{\odot}$/yr, which is consistent with previous results (\citealt{2010ApJ...716L.103L}, \citealt{2010A&A...515A..73S}).



\begin{figure}
   \centering
           \includegraphics[width=9.5cm]{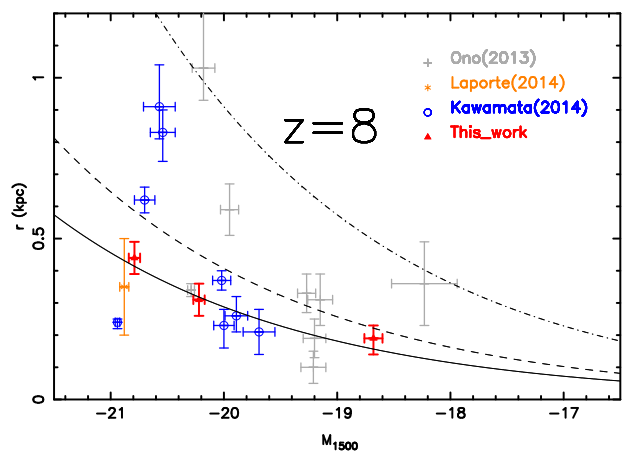}
      \caption{\label{fig.size} Position of the 3 best $z$$\sim$8 candidates selected in this study in a size-luminosity diagram (red dots). For comparison purposes, we plotted the size of objects selected using the first Frontier Fields dataset (\citealt{2014arXiv1410.1535K}, \citealt{2014A&A...562L...8L}) and results from the HUDF \citep{2013ApJ...777..155O}. Lines display the size-luminosity relation assuming several star formation rate densities: 10 (black), 5 (dashed), and 1 (dash-dotted) M$_{\odot}$yr$^{-1}$kpc$^{-2}$ }
   \end{figure}

We computed the size of our targets using the galaxy-fitting software \textit{GALFIT} \citep{2002AJ....124..266P} assuming that they are well fitted by a Sersic profile, as previously shown, for example, in \citet{2010ApJ...709L..21O}. We fixed the Sersic index to n=1 following the method described in \citet{2014arXiv1410.1535K} and the discussion on the Sersic index published in that paper and in \citet{2013ApJ...777..155O}. The amplification was taken into account by dividing the radius of our target by the amplification factor estimated in Section \ref{sec.magnification}. For comparison purposes, we also checked that these results are consistent with the size computed from the  SExtractor half-light radius corrected for PSF-broadening and magnification, in a similar manner as in \citet{2014A&A...562L...8L}. The sizes of our targets range from 0.2 (for the faintest) to 0.4 kpc (for the brightest). We also plotted the sizes of our targets as a function of their luminosities (excluding MACS0416\_Y3 that is more likely a $z\sim$9 object).The position of the $z$$\sim$8 candidates selected behind MACS0416 is consistent with previous findings, in the sense that they are not located outside of the region covered by previous studies (see Figure \ref{fig.size}).  As already noted by other groups, the size of $z$$\sim$8 galaxies seems to be correlated with the UV luminosity. We used Eq. 4 of \citet{2013ApJ...777..155O} to show that the evolution of the size of the candidates selected in this study involves a constant star formation rate density of $\Sigma_{SFR}$=10 M$_{\odot}$yr$^{-1}$kpc$^{-2}$.  Table \ref{tab.z8_prop} presents the SED-fitting results and the properties we can deduce from their photometry. 

%
%

\section{Contaminants}
\label{sec.contaminants}
Photometric samples are subject to contamination by low-$z$ objects, and a key issue to avoid biasing the statistical study of very high-redshift galaxy properties is to identify genuine high-redshift objects from interlopers. The bulk of contaminants to $z$$\sim$8 samples are supernovae, low-mass stars,  emission-line galaxies, and photometric scatter of low-$z$ objects. The observations of MACS0416$-$2403 by {\it HST} span several months, allowing the removal of transient object (supernovae) in the final mosaics. Moreover, all of the sources in this $z$$\sim$8 sample are resolved, effectively ruling out a low-mass star origin as well.
 
\begin{figure*}
   \centering
           \includegraphics[width=15cm]{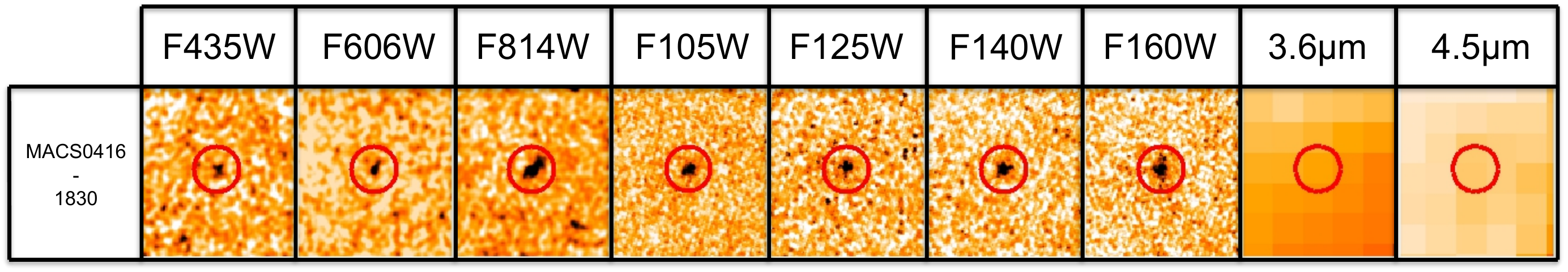}
      \caption{\label{fig.Bradley_stamps}Thumbnail images of MACS0416$-$1830 selected as $z$$\sim$8 candidate using CLASH data (\citealt{2012ApJS..199...25P} and \citealt{2014ApJ...792...76B}). ACS data from the HFFs are deeper than those provided by the CLASH survey, demonstrating that this object is a $z$$\sim$2 interloper. We used the SED of this source as a basis to estimate the contamination rate of the sample of $z$$\sim$8 galaxies presented in this paper.}
   \end{figure*}

\begin{figure*}
   \centering
           \includegraphics[width=15cm]{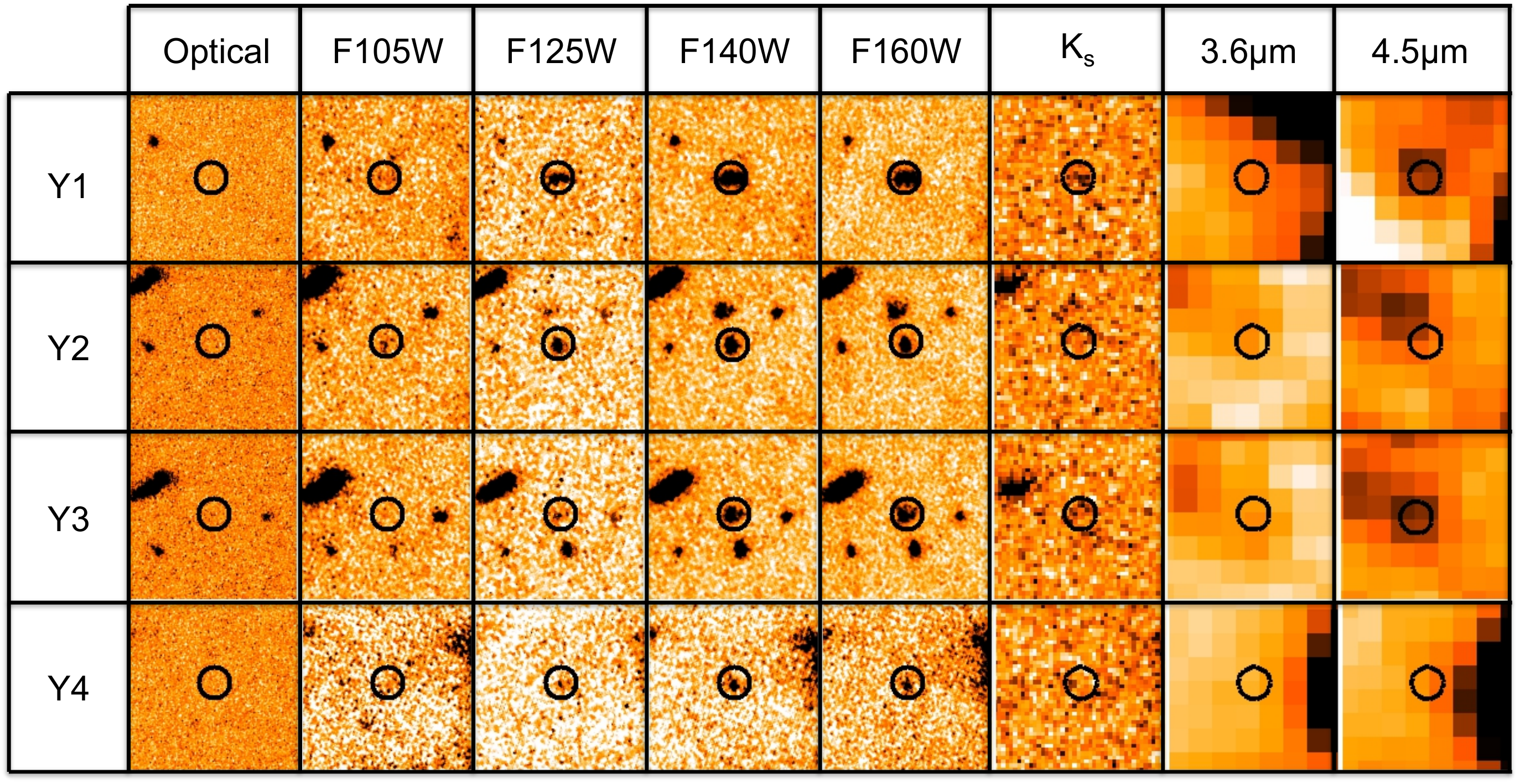}
      \caption{\label{fig.z8_stamps}Thumbnail images of the four objects selected using HFF data of cluster MACS0416$-$2403. Each stamp is 3''$\times$3'', and the position of the candidate is displayed as a red circle of 0\farcs4 radius. The first column shows the stacked image of HST optical bands (F435W, F606W and F814W) at the position of the $z$$\sim$8 candidate. }
   \end{figure*}

Thanks to the depth of the HFF optical images of MACS0416$-$2403, we were able to identify a previous $z$$\sim$8 candidate selected from the CLASH survey \citep{2012ApJS..199...25P} as a low-$z$ interloper. This object, named MACS0416-1830 \citep{2014ApJ...792...76B}, follows the $z$$\sim$8 color criteria defined in Section \ref{selection_criteria}, but is detected in all ACS bands (Figure \ref{fig.Bradley_stamps}) with a F606W-F125W break of $\approx$1.5 mag, implying it is unlikely to be at such high redshift. The best SED fit is found by following the method described in Sect. \ref{sec.sed-fitting} for a redshift of $z$$=$1.77$^{+ 0.80}_{-1.77}$ and a small reddening value, $A_v$=0.20. We used its best-fit SED to estimate the contamination rate in the sample of $z$$\sim$8 galaxies presented in this study. Because of the large break required by our critieria between F814W and F125W, none of our selected objects have SEDs similar to this kind of contaminant. 

We also applied the same method to all of the $z$$\sim$8 candidates previously published using HFF data. \citet{2014arXiv1409.0512A} selected 8 $z$$\sim$8 candidates, of which two sources (1060 and 6593) might be interlopers identical to the MACS0416$-$1830 interloper. Moreover, two objects from the $z$$\sim$8 sample published by \citet{2014arXiv1402.6743Z} display SEDs similar to this interloper (ZD1 and ZD10). Therefore, we estimate that 20-25\% of the $z$$\sim$8 sample selected from deep surveys might be faint interlopers. The identification of a faint interloper like this one provides a warning against over interpreting the likelihood of the faintest LB candidates, where the optical data are not sufficient to probe such strong Balmer breaks. This criterion effectively limits the selection of candidates in MACS0416$-$2403 to sources with F125W $<$28.2.

%
%

\section{Luminosity function at $z\sim$8}
\label{sec.LF}
The evolution of galaxies at very high-redshift is still unclear. Recent studies suggest a strong dimming in the number of galaxies beyond $z$$\sim$7 \citep{2014arXiv1403.4295B}. In the following we compute the UV LF at $z$$\sim$8, taking advantage of the previous SED-fitting work.
\subsection{Method}
We adapted the method described in \citet{2002A&A...395..443B} incorporating the previously assessed redshift probability distribution, $P(z)$, for each source. We summarize below the steps we followed:
\begin{itemize}
\item A random probability (called $p_a$) is chosen, and a redshift ($z_a$) is assigned using the cumulative $P(z)$ computed for each object, i.e., $p_a$=$P_{cum}(z_a)$.
\item The UV luminosity is computed from the redshift $z_a$ assigned at the previous step and the SED of each source.
\item The previous steps are repeated $N$ times to obtain a sample with $N$ times the size of the original sample, but with the same redshift distribution.
\item Each simulated object is distributed into bins of $M_{UV}$, and each bin is divided by the number of simulations, $N$,  and the comoving volume explored. Then each density is corrected for completeness. In the case of a cluster field, we computed the effective volume covered by this survey in a similar manner as in \citet{2014arXiv1405.0011C} and using a mask of the bright objects in the cluster core. 
\item Error bars include statistical uncertainties and cosmic variance computed using the integration of a two-point correlation function over the volume explored by our survey \citep{2008ApJ...676..767T}
\end{itemize}
The adopted number of iterations, $N$, is chosen so as to not affect the end result; in our case, we used several iterations number ranging from 50 to 10 000 and showed that the results does not change for $N$  above 1000 iterations.

\subsection{Results}
We computed the effective volume covered by our survey following the method described in \citet{2014arXiv1405.0011C}, using the amplification map provided by the CATS group and by masking the area where bright objects are on the detection image. Three densities can be computed at $M_{1500}$=-21.00, -20-00 and -19.00 with error bars including Poison errors and Cosmic Variance (see Table\ref{tab.LF_densities}) . 

\begin{figure*}
   \centering
           \includegraphics[width=15cm]{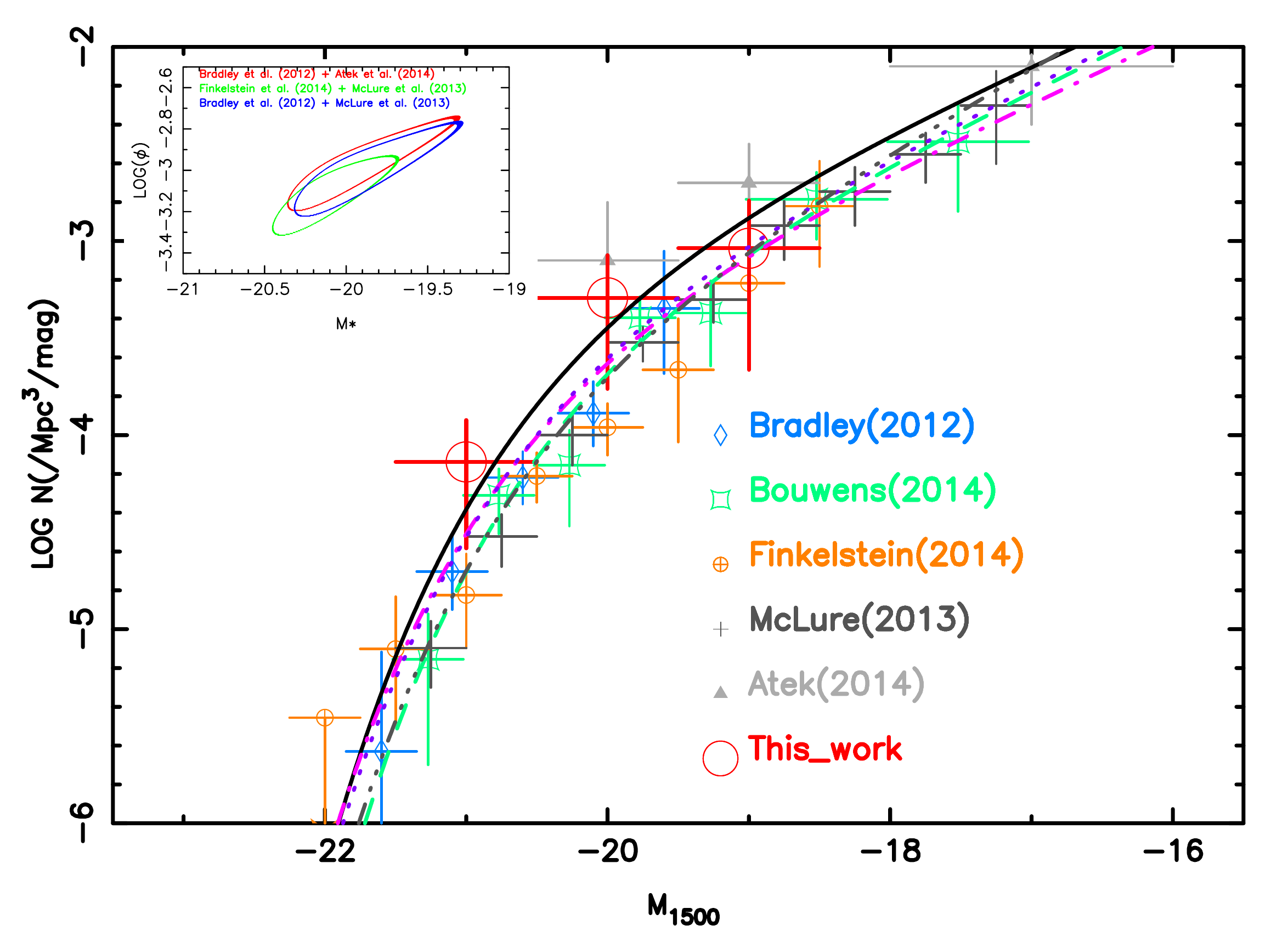}
      \caption{\label{fig.LF} UV luminosity function at $z$$\sim$8 showing the number densities computed from an MC method based on SED-fitting results (in red). The best fit of the Schechter function is shown as the black line. For comparison purpose, we over plotted densities from \citet{2014arXiv1409.0512A}, \citet{2012ApJ...760..108B}, \citet{2014arXiv1403.4295B}, \citet{2013MNRAS.432.2696M}, and \citet{2014arXiv1410.5439F}. The green dashed line is the LF published by \citet{2014arXiv1403.4295B}, the gray line the parameterization from \citet{2013MNRAS.432.2696M}, the pink line is from \citet{2014arXiv1408.6903I}, and the magenta line from \citet{2014ApJ...786...57S}. The left upper panels displays the 1$\sigma$ confidence interval on the fit of the Schechter function using 3 combinations of numbers densities: Bradley et al. (2012) and Atek et al. (2014) in red, Finkelstein et al. (2014) and McLure et al. (2013) in green, and Bradley et al. (2012) and McLure et al. (2013) in blue . }
   \end{figure*}

\begin{figure*}
   \centering
           \includegraphics[width=19cm]{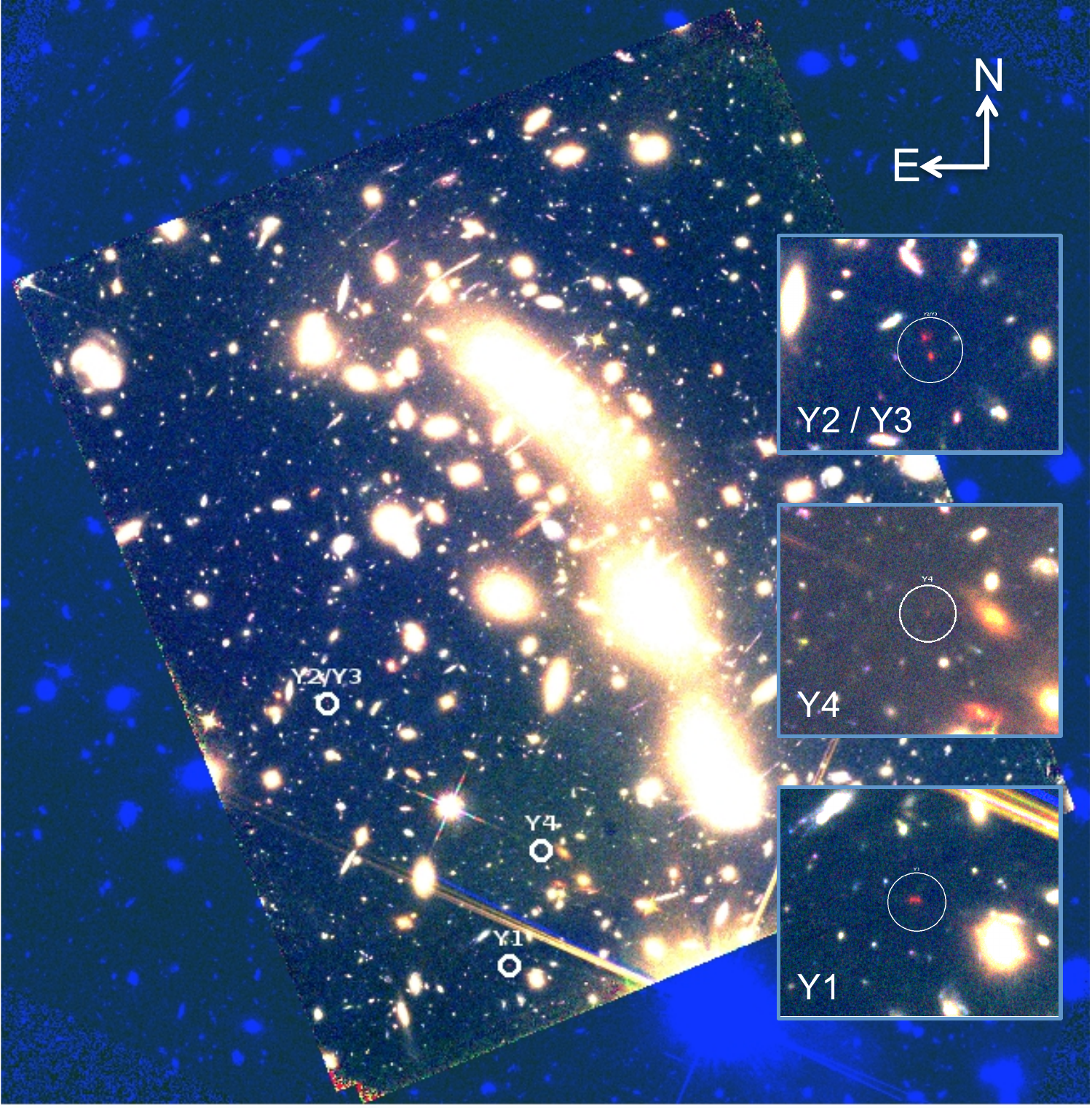}
      \caption{\label{fig.FoV} RGB images (blue: F435W, F606W and F814W, green: F105W , red: F160W) showing the position of the 4 candidates discussed in this paper.  North is up and east to the left.  }
   \end{figure*}

We adopted a Schechter parameterization  \citep{1976ApJ...203..297S} and fitted the shape of the UV LF at $z$$\sim$8 using a $\chi^2$ minimization method. To show the influence of our densities on the shape of the LF, we computed the three Schechter parameters using the density at the brightest luminosities published by \citet{2014ApJ...792...76B} and at the faintest luminosities using results from \citet{2014arXiv1409.0512A}. Error bars on each parameter are given by the 1$\sigma$ confidence interval (see Figure \ref{fig.LF}). 
We also used others combinations of results to fit the shape of the UV luminosity function at $z\sim$8 to confirm the range of parameters we found.  For each set of number densities, the 1$\sigma$ confidence intervals agree well. Table \ref{tab.fit_LF} shows that our parameterization is consistent with previous results in this range of redshift, and Fig. \ref{fig.LF} displays the shape of the UV LF at $z$$\sim$8 we found using the sample presented in this study.


\begin{table}
\caption{Fit of the UV luminosity function at $z$$\sim$8}             
\label{tab.fit_LF}      
\centering                          
\begin{tabular}{p{1.7cm} | c c c c | }        
\hline\hline                 
References 				&	M$^{\star}$			& $\alpha$				& $\Phi^{\star}$		\\
						&	[mag]				&						& $\times$10$^{-3}$Mpc$^{-3}$ \\
						&						&						&					\\  \hline
This work 					&	-20.07$\pm$0.52		& -1.80$\pm$0.31			& 0.94$^{+0.64}_{-0.65}$ \\
						&						&						&					\\  					
\citet{2014arXiv1403.4295B}	&	-19.97$\pm$0.34		& -1.86$\pm$0.27			& 0.64$^{+0.65}_{-0.32}$ \\
						&						&						&					\\  
\citet{2014arXiv1408.6903I}	&	-20.20$^{+0.20}_{-0.50}$	& -1.83$^{+0.25}_{-0.28}$ 	& 0.50$^{+0.91}_{-0.21}$ \\
						&						&						&					\\  
\citet{2014ApJ...786...57S}	&	-20.15$^{+0.29}_{-0.38}$	& -1.87$\pm$0.26 			& 0.57$^{+0.45}_{-0.31}$ \\
						&						&						&					\\  						
\citet{2013MNRAS.432.2696M}	&	-20.12$^{+0.37}_{-0.48}$	& -2.02$^{+0.22}_{-0.23}$ 	& 0.47$^{+0.67}_{-0.40}$ \\
						&						&						&					\\  
\hline
\hline                                   
\end{tabular}
\end{table}
%
%


\begin{table}
\caption{Number densities at $z$$\sim$8}             
\label{tab.LF_densities}      
\centering                          
\begin{tabular}{cc}        
\hline\hline                 
M$_{1500}$ 				&	$\Phi$(M$_{1500}$)			\\
\hspace{0.1cm} [mag]					&	$\times$10$^{-4}$ [/Mpc$^3$/mag]			\\\hline
						&							\\  
-21.00$\pm$0.50			&	0.73$\pm$0.47			 \\
-20.00$\pm$0.50			&	5.09$\pm$3.36						\\  					
-19.00$\pm$0.50			&	9.18$\pm$7.01			 \\  
\hline
\hline                                   
\end{tabular}
\end{table}
%
%

\section{Discussion}
\label{sec.discussion}
MACS0416$-$2403 is the first HFF cluster observed that overlaps with previous {\it HST} observations performed by the CLASH survey \citep{2012ApJS..199...25P}. 
However, the depth of the HFF data is at least 1 magnitude deeper than those from the CLASH survey and hence allow a check on previous conclusions. \citet{2014ApJ...792...76B} published only one object that appeared as a $z$$\sim$8 galaxy in CLASH data, but this object is now clearly detected at optical wavelengths in the HFF survey (Figure \ref{fig.Bradley_stamps}), confirming that there are no true high-z galaxies brighter than F125W=26.5 in MACS0416$-$2403, as shown by \citet{2012arXiv1211.2230B}. The brightest galaxy selected using the HFF, MACS0416\_Y1, lies outside the field of view covered by CLASH, which explains why it was not included in previous samples. 

We checked the consistency of our sample size with previous results by estimating the expected number of $z$$\sim$8 galaxies with $m_{F125W}\leq$28.2 in the field of view covered by this survey and corrected for magnification. We used the evolution equations of the UV LF from previous HST Legacy fields \citep{2014arXiv1403.4295B}, including error bars on Schechter parameters and cosmic variance. We found that 1.6$^{+4.5}_{-1.2}$ $z\sim$8 galaxies should be included in this sample, showing that our sample is fully consistent with previous observational results. We also used a set of previous parameterization of the UV LF at $z$$\sim$8 to confirm this expectation:  \citet{2014arXiv1408.6903I} (N=1.6$^{+4.1}_{-0.7}$), \citet{2013MNRAS.432.2696M} (N=1.5$^{+7.8}_{-1.3}$), \citet{2013ApJ...768..196S} (N=2.00$^{+6.5}_{-1.4}$), and \citet{2012ApJ...759..135O} (N=1.5$^{+8.0}_{-1.2}$). All these expectations are consistent and demonstrate that our sample agree well with them.


\section{Conclusions}
We described the search for $z$$\sim$8 objects behind MACS0416$-$2403, the second cluster observed in the framework of the HFF project. We combined data collected with \textit{HST}, \textit{VLT} and \textit{Spitzer} to have a better coverage of the SED from $\sim$0.4$\mu$m to $\sim$5$\mu$m. Four (4) objects, with m$_{F125W}$ ranging from 26.3 to 27.5, selected following the LBG method,  appear as good $z$$\sim$8 candidates. Moreover, two of them display a break in the Spitzer data, suggesting possible contamination by [OIII] and H$\beta$ emission lines at $z$$\sim$8 and/or the detection of the Balmer break at very high-redshift \citep{2014ApJ...784...58S}. The SED-fitting analysis demonstrated that all of the objects in the sample prefer high-$z$ solutions, without any reliable secondary solution at low-$z$ (i.e., with P($z$)$>$5\%). 
The SFR of our candidates agree well with expectations and previous observational results. We also computed the size of each candidate using a galaxy-fitting software and showed that all objects in our sample have a size ranging from 0.2 to 0.5 kpc and that there clearly is an evolution of the size as a function of luminosity.

We computed number densities of objects using MC simulations based on the redshift probability distribution computed during the SED-fitting analysis. We corrected the values we found for cosmic variance and fitted the UV LF using previous results in the same range of redshift. The Schechter parameters we found (M$^{\star}$=-20.07$\pm$0.52,   $\Phi^{\star}$=0.94$^{+0.64}_{-0.65}$ $\times$10$^{-3}$Mpc$^{-3}$ and $\alpha$=-1.80$\pm$0.31) confirm the trend observed for a few years at $z$$\sim$8. However, results should be taken with caution because of the large error bars on each parameter.

Thanks to the depth of the HFF data, we identified one previous $z$$\sim$8 candidate selected from the CLASH survey as a mid-$z$ interloper, with a best photo-$z$ of $z=$1.77$^{+ 0.80}_{-1.77}$. Using the best-fit SED for that object as a template, we estimated the contamination rate by this kind of interlopers in our sample and previous samples built using HFF data. By design, none of our $z$$\sim$8 candidates selected behind MACS0416$-$2403 displays an SED similar to this kind of contaminant, since we required a break of at least 2.5 magnitudes between optical and NIR data. We estimated that $\sim$20-25\% of objects selected using the HFF data on Abell 2744 are consistent with such faint mid-$z$ interlopers.  

All of these conclusions are based on a photometric analysis and ultimately need spectroscopic confirmation. Even if the detection of Lyman $\alpha$ at very high-redshift is challenging (e.g., \citealt{2014MNRAS.443.2831C}), the continuum of the brightest objects should be observable with extremely deep observations using current facilities, such as  KMOS/VLT \citep{2004SPIE.5492.1179S}, FLAMINGOS-2/Gemini \citep{2006SPIE.6269E..17E} and NIRSPEC/Keck \citep{1998SPIE.3354..566M}.

\begin{acknowledgements}
The authors thank the anonymous referee for useful comments that improved the quality of the paper, Mathilde Jauzac and Johan Richard for providing amplification maps for the cluster. We acknowledge support from CONICYT-Chile grants Basal-CATA PFB-06/2007 (NL, FEB, SK), Gemini-CONICYT \#32120003 (NL), "EMBIGGEN" Anillo ACT1101 (FEB), FONDECYT 1141218 (FEB), and Project IC120009 ``Millennium Institute of Astrophysics (MAS)'' of the Iniciativa Cient\'{\i}fica Milenio del Ministerio de Econom\'{\i}a, Fomento y Turismo (FEB), the Spanish Ministry of Economy and Competitiveness (MINECO) under the 2011 Severo Ochoa Program MINECO SEV-2011-0187 (AS), the French Agence Nationale de la Recherche bearing the reference ANR-09-BLAN-0234 (RP, DB). Based on data obtained from the ESO Science Archive Facility under request number Laporte112957. This work is based on observations made with the NASA/ESA Hubble Space Telescope, obtained at the Space Telescope Science Institute (STScI), which is operated by the Association of Universities for Research in Astronomy, Inc., under NASA contract NAS 5-26555. The HST image mosaics were produced by the Frontier Fields Science Data Products Team at STScI. This work is based in part on observations made with the Spitzer Space Telescope, which is operated by the Jet Propulsion Laboratory, California Institute of Technology under a contract with NASA.

\end{acknowledgements}

\bibliographystyle{aa}  
\bibliography{laporte_macs0416.bib} 

%
%
%
%
%
%

\end{document}